\documentclass[ ]{article}

\usepackage{url}
\usepackage{xurl}
\usepackage{times}
\usepackage{ragged2e}
\usepackage{graphicx} 
\usepackage{tabularx}
\usepackage{hyperref}
\usepackage{amsmath}
\usepackage{relsize}
\usepackage[gen]{eurosym}
\usepackage{soul}
\usepackage{cancel}
\usepackage{fix-cm}
\usepackage{microtype}

\makeatletter
\long\def\@makecaption#1#2{%
  \vskip\abovecaptionskip
  \sbox\@tempboxa{\textit{#1: #2}}%
  \ifdim \wd\@tempboxa >\hsize
    \textit{#1: #2}\par
  \else
    \global \@minipagefalse
    \hb@xt@\hsize{\hfil\box\@tempboxa\hfil}%
  \fi
  \vskip\belowcaptionskip
}
\makeatother

\begin{document}

\title{Assessing Honey Bee Colony Health Using Temperature Time Series }

\author{
Karina Arias-Calluari$^{1,2}$,
Theotime Colin$^{3}$,
Tanya Latty$^{2}$,\\
Mary Myerscough$^{1}$ and
Eduardo G. Altmann$^{1}$
}
\date{}
\maketitle
\begin{center}
$^{1}$School of Mathematics and Statistics, The University of Sydney, Sydney, Australia\\
$^{2}$School of Life and Environmental Sciences, The University of Sydney, Sydney, Australia\\
$^{3}$School of Natural Sciences, Macquarie University, Sydney, Australia
\end{center}

\begin{abstract}

Honey bees face an increasing number of stressors that disrupt the natural behaviour of colonies and, in extreme cases, can lead to their collapse.  Quantifying the status and resilience of colonies is essential to measure the impact of stressors and to identify colonies at risk. In this manuscript, we present and apply new methodologies to efficiently diagnose the status of a honey bee colony from widely available time series of hive and environmental temperature. Healthy hives have a remarkable ability to control temperature near the brood area. Our method exploits this fact and quantifies the status of a hive by measuring how resilient they are to extreme environmental temperatures, which act as natural stressors. Analysing 22 hives during different times of the year, including 3 hives that collapsed, we find the statistical signatures of stress that reveal whether honeybees are doing well or are at risk of failure. Based on these analyses, we propose a simple scale of hive status (stable, warning, and collapse) that can be determined based on a few temperature measurements. Our approach offers a lower-cost and practical bee-monitoring solution, providing a non-invasive way to track hive conditions and trigger interventions to save the hives from collapse.
\end{abstract}
\maketitle
\vspace{1.0cm}
\section{Introduction}
Globally, the annual market of insect pollination has been estimated at around US\$212 to US\$577 billion \cite{vanEngelsdorp2010S80,pollinators2016pollinators}. Honey bees (\textit{Apis mellifera}) are widely recognised as the most significant pollinators of many crops \cite{vanEngelsdorp2010S80,braga2020method,kane2021honey,martinez2022migratory}, and due to their ecological significance, honey bee populations directly impact biodiversity worldwide \cite{hung2018worldwide}. The manageability and adaptability characteristics of honey bees make them highly effective as commercial pollinators, resulting in a large number of globally distributed managed hives \cite{phiri2022uptrend}. This, in turn, has enhanced the impact of honey bees on food supply, global food security, and has major financial implications worldwide \cite{kane2021honey,calderone2012insect,ramanujan2012insect}. For instance, in the USA, it has been estimated since 2016 that honey bees pollinate crops worth US\$15 billion annually \cite{USA1,USA2}. Since 2015, bees across the European Union have contributed at least EUR \euro{22} billion annually to the agriculture industry \cite{EU1,EU2}. In Australia, the honey bee industry has been worth approximately AUD\$14 billion annually since 2022 \cite{Australia}.\par

Unfortunately, maintaining the health of honey bee colonies is increasingly challenging. Colony failures have been attributed to multifactorial causes that include poor nutrition\cite{goulson2015bee}, adverse weather \cite{oldroyd2007s}, pesticide exposure \cite{pettis2013crop,colin2019traces}, and parasites like varroa mites \cite{degrandi2016population,alma991031528020305106,martin2020varroa}, resulting in diverse types of colony failure such as starvation \cite{steinhauer2014national}, queen failure \cite{steinhauer2014national}, infestations or Colony Collapse Disorder syndrome \cite{vanengelsdorp2009colony,cox2007metagenomic}. As the average global temperature continuously increases \cite{warming_this_year}, resulting in changes in local climate conditions \cite{bergstrom2021combating,keith2017ecosystem,Australia2}, the escalation of threats and stressors faced by honey bees underscores the importance of identifying stress indicators early and facilitating timely interventions to enhance hive survival.\par

Extensive efforts have been made to unravel the causes of hive failures and establish warning signs of stress to prevent hive collapse. The best-known approaches require direct manipulation of the hive, the most common being inspection procedures \cite{braga2020method,tarpy2013idiopathic} and toxicological approaches \cite{hendriksma2011honey}. Unfortunately, these direct approaches can destabilise hives if not performed correctly, and they are often restricted by weather conditions to avoid unnecessary risks when opening the hive. On the other hand, more sophisticated mathematical models can provide an approximate representation or estimation of the emerging behaviour of population dynamics of honey bee colonies \cite{perry2015rapid,khoury2013modelling}. Although these models significantly enhance the understanding of the underlying mechanism of colonies transition from a state of apparent health to failure, how to implement them in daily hive management is unclear. Other strategies focus on estimating colony populations using empirical data extracted from the hives. These integrated approaches may reduce the need for frequent physical hive manipulations by pairing data, typically collected via sensors \cite{cecchi2020smart,ferrari2008monitoring}, with current models. 
For instance, models may pair foraging activity to weather condition \cite{clarke2018predictive} or weight of the hive \cite{arias2023modelling}, and in-hive bee population with the brood nest temperature \cite{seeley1985survival,becher2010brood}. Thence, an approximation of the colony size can be derived from empirically collected hive data to assess whether the colony population has stabilised during the evaluated period. However, the length and precision of the datasets collected directly influence the final result, adding costs when aiming for higher precision. More recent approaches involve using machine learning algorithms to construct decision trees based on extensive data collections \cite{sharif2022monitoring}, and other technologies attempt to utilise intelligent monitoring systems for image processing to determine the population of foraging bees\cite{tashakkori2015image} or detect infestations \cite{braga2021intelligent}, but in most cases are expensive and require extensive maintenance. 

In this manuscript, we introduce a new methodology for monitoring the hive's status at a lower cost and with minimal data compared to alternative methods. We propose to assess the status of the hive by quantifying the ability of the bees to thermoregulate the hive's internal temperature when exposed to extreme environmental temperatures, which act as natural stressors of the hives. High-resolution measurements of environmental and hive temperatures are cheaply available and used as inputs of our methodology to quantify the status of the hive. Applying our methodology to 22 independent hives at different times of the year, we succeed in classifying healthy hives from hives that are struggling and approaching collapse.
\begin{figure*}[hbt!]
\includegraphics[scale=.52,trim=0cm 0cm 0cm 0cm, angle =0 ]{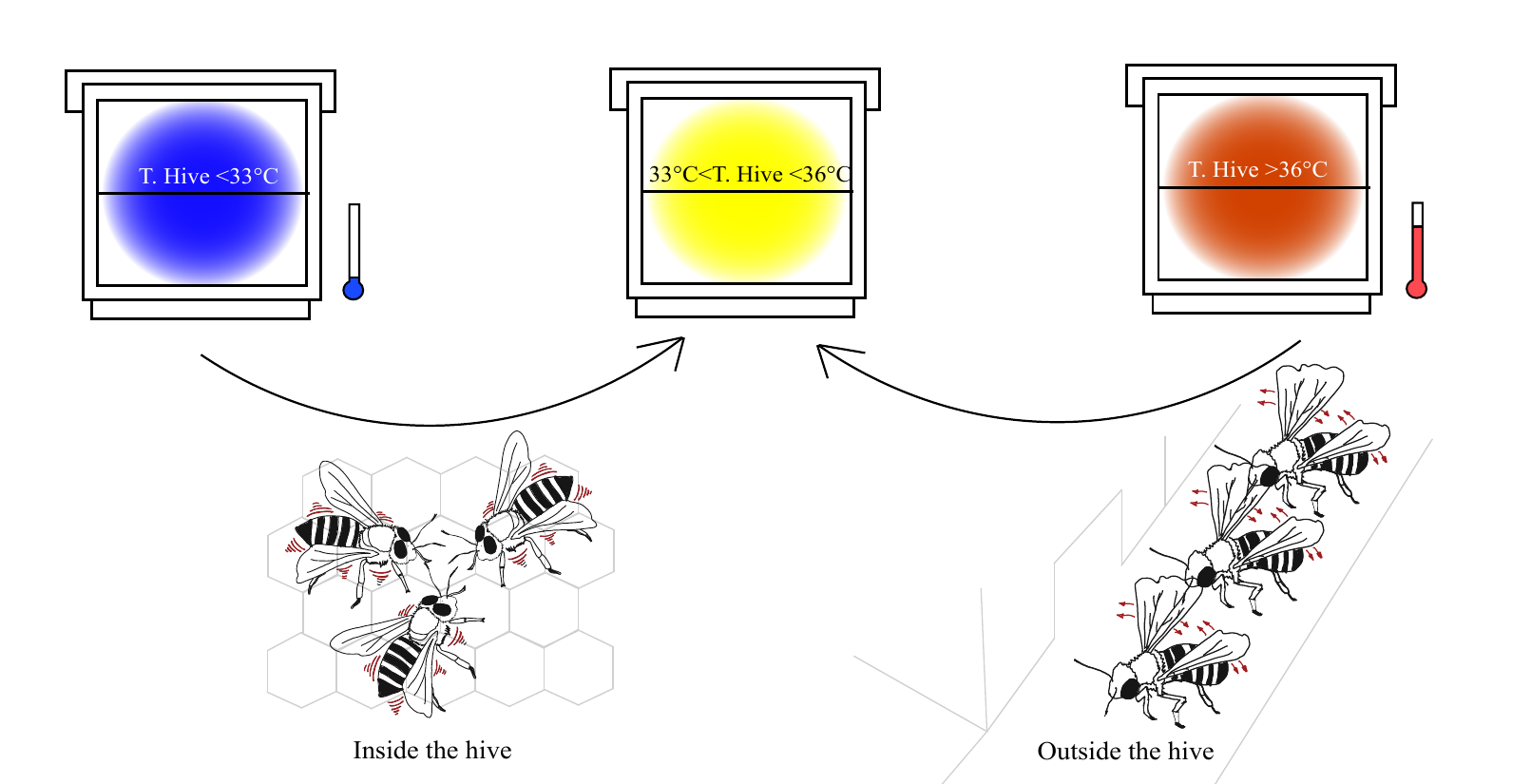}
\caption{Illustration of hive bee's activities when dealing with temperatures outside the range of 33$^{\circ}$C to 36$^{\circ}$C. In situations of lower environmental temperatures, when the hive's temperature is much lower than 33$^{\circ}$C, especially during cold winters, bees cluster together and generate metabolic heat by contracting and relaxing their flight muscles, a process known as `shivering'. The result of this vibration is an increase in their thoracic temperature, which allows them to thermally control their hive. On the other hand, when the hive's temperature is higher than 36$^{\circ}$C, bees ventilate the hive by fanning the hot air out or using evaporative cooling mechanisms \cite{stabentheiner2003endothermic}.}
\label{fig.illustration}
\end{figure*}

\section{Methodology}
\label{Sec:Methodology}
\subsection{General idea}
The starting point for our methodology is the observation that maintaining brood nest temperature between 33°C and 36°C is crucial for normal development of larvae and pupae as well as other in-hive activities~\cite{cook2016rapidly,stabentheiner2010honeybee,ocko2014collective}. Honey bees have developed remarkable strategies to collectively thermoregulate hive temperatures \cite{peters2019collective}, as illustrated in figure~\ref{fig.illustration}. These strategies require a sufficient number of healthy bees to act in coordination \cite{jhawar2023honeybees}. This motivates the key assumption of our methodology: the status of a hive can be quantified by its ability to bring the hive temperature $T_H$ to the desired temperature $T_d \in [33^\circ C, 36^\circ C]$.

\begin{figure*} [hbt!]
\includegraphics[scale=0.6,trim=-2cm 1cm 0cm 0cm, angle =0 ]{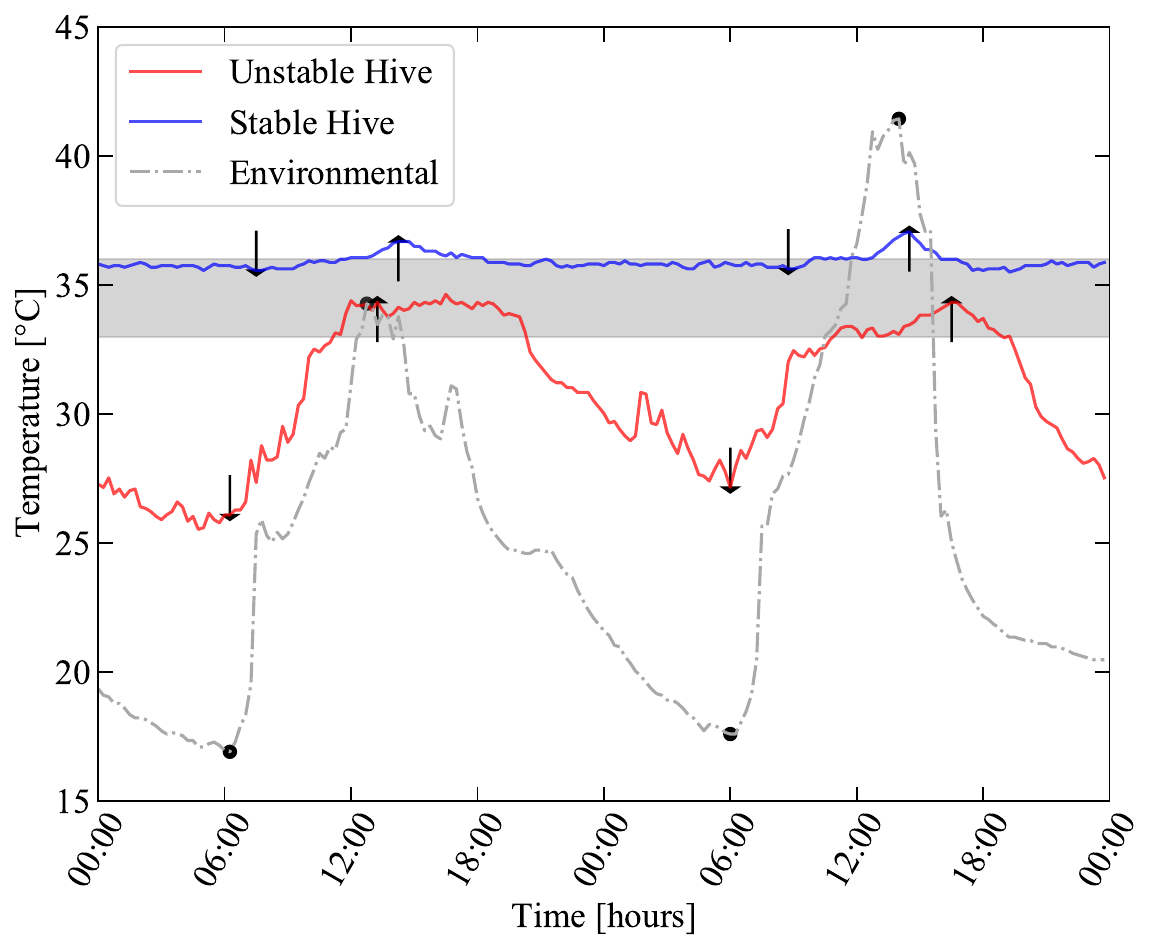}
\caption{Low and high environmental temperatures affect the hive's core temperature. The plot shows the variation in the temperature of two hives  (coloured lines) subject to the same environmental conditions (dot-dashed line) in a period of 48h. Extremes in the environmental temperature $T_E$ (denoted by $\bullet$) lead to peaks (top arrow $\uparrow$) and valleys (bottom arrow $\downarrow$) in the core temperature of both hives. These variations are large in an unstable hive (red line) and small in a stable one (blue line), confirming the resilience of hive  to environmental temperature change as the result of hive bee activities. The shade regime reflects the desired hive temperature.}
\label{low_and_high_temperatures}
\end{figure*}

An important factor to be considered in this analysis is the environmental temperature $T_E$. For $T_E \approx T_d$, bees are not challenged, and little information about the status of the hive can be obtained. In contrast, for $T_E \gg T_d$ or $T_E \ll T_d$, the external temperature can be seen as a stressor and only well-prepared hives will be able to bring $T_H$ close to $T_d$. Therefore, we build our methodology by quantifying the interplay between $T_H$ and $T_E$, more precisely, quantifying the extent into which $T_H$ is susceptible to variations in $T_E$. Stable hives are expected to show small susceptibility while unstable hives display a larger susceptibility. Figure~\ref{low_and_high_temperatures} confirms this expectation and illustrates the main effect our methodology explores. For two hives subjected to the same $T_E$, we see an evident difference in the variation of their $T_H$. Importantly, we observe a time delay $\tau$ between the variations in $T_E$ and $T_H$, reflecting the thermal inertia of the hive and the reaction of the bees.

\subsection{Quantitative methodology}
For a given hive, we assume that the hive temperature $T_H$ and the environmental temperature $T_E$ are available at a selection of times $t=0, 1, \ldots, t_{max}$. Our model is that $T_H=T_H(t+\tau)$, where $\tau > 0 $, depends linearly on $T_E = T_E(t)$ and we thus write
\begin{equation}
T_H=(T_E - (T_d - \Delta T))m +T_d,
\label{Eq:model}
\end{equation}
where $0 \leq m\leq 1$ is the slope of the linear relationship, $T_d$ the desired temperature, and $\Delta T$ the temperature change due to intrinsic activities of the bees and their metabolic heat. Together, $T_d$ and $\Delta T$ determine the intercept point of the linear dependence of $T_H$ on $T_E$ at $T_E=0$. Overall, equation~(\ref{Eq:model}) satisfies $T_H=T_d$ for $m=0$ (perfectly efficient hive), $T_H = T_E$ for $m=1$ (totally inefficient or empty hive with $\Delta T$=0), and $T_H=T_d$ for any $m$ if $T_E = T_d - \Delta T$. This last point shows that $\Delta T$ can be interpreted as the gap between the most comfortable $T_E$ for the hive and $T_d$. Indeed, we systematically find $\Delta T > 0$, a reflection of the fact that bees inherently perform heat-generating activities which, in the absence of temperature-controlling activities, would lead to $T_H>T_d$ for $T_E=T_d$. 

The description above suggests that we consider two quantitative indicators of the status of the hive, $\Delta T$ and $\Pi \equiv - \log m$. The motivation for taking $\Pi$ as a non-linear transformation of $m$ is that it uniquely maps $m \in (0,1]$ into $\Pi \in [0,\infty)$, providing a better distinction between the stable or efficient (small $m$ or large $\Pi$) and unstable or inefficient (large $m$ or small $\Pi$) hives. Moreover, both measures can be taken as comparable measures of performance because they have the same range $\Delta T \ge 0$ and $\Pi \ge0$, with $\Delta T = 0$ and $\Pi = 0$ for empty hives (no bee activities), and they increase as hive performance gets better. In practice, we observe that healthy hives typically show $\Delta T \approx 10$ and $\Pi \approx 3$, so values higher than these are considered irrelevant. We compute $\Delta T$ and $\Pi$ using two different methods, as described below.

\subsubsection{Method 1: extreme temperatures}
We extract the maximum and minimum environmental temperatures daily from $T_E(t)$ and analyse the hive's temperature response $H(t+ \tau)$ for $\tau$ up to 2 hours after either the minimum or maximum environmental temperature was recorded. Then, we compute $\Pi$ and $\Delta T$ over a specific time window $t_w$ from equation~(\ref{Eq:model}) directly.\par
\begin{figure*} [hbt!]
\includegraphics[scale=0.47,trim=0cm 0cm 0cm 0cm, angle =0 ]{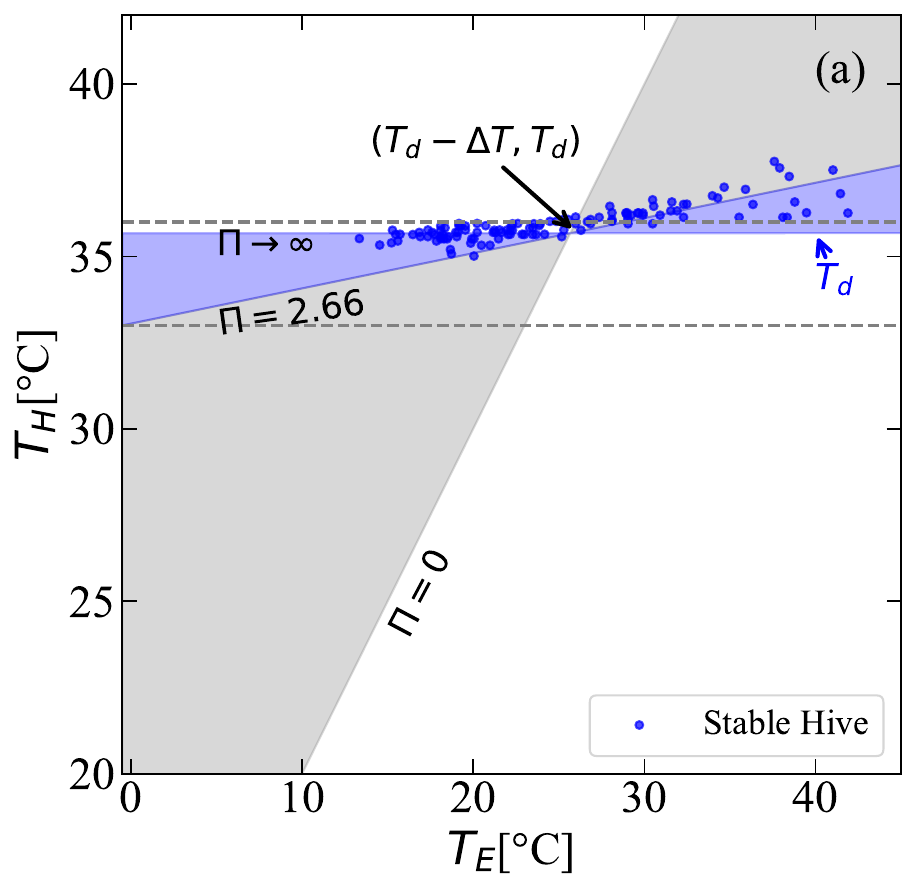}
\includegraphics[scale=0.47,trim=0cm 0cm 0cm 0cm, angle =0 ]{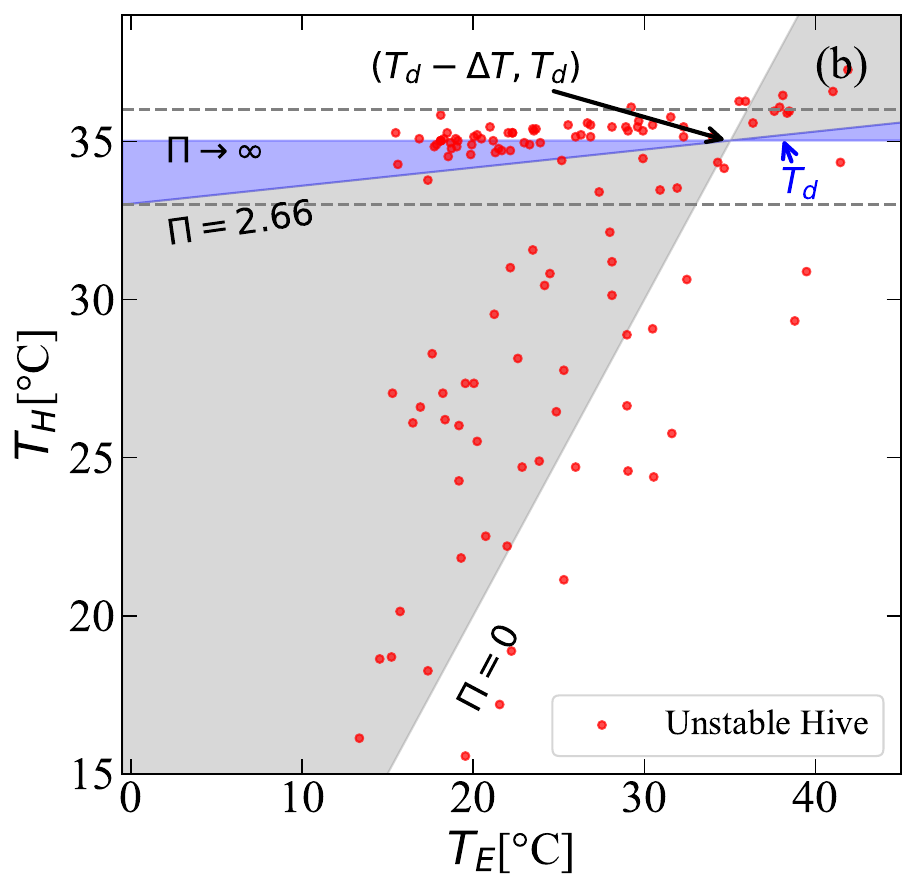}
\caption{Illustration of the minimum and maximum environmental temperature registered in situ against the core's hive temperature. For the (a) stable hive and for the (b) unstable hive, both of them are under the same temporal and spatial conditions. The unstable hive loses control of the temperature gradually ($\Pi< \,2.66$) until it eventually collapses ($\Pi= \,0$).}
\label{Th_Te}
\end{figure*}

In subfigures \ref{Th_Te}~(a-b), we depict the possible responses of a hive where a threshold of $\Pi=2.66$ corresponds to a hive temperature within the desired range, given an environmental temperature difference up to 42°C. From $0 \leq \Pi < 2.66$, a grey zone represents a transition area, lying outside the desired range and extending toward collapse ($\Pi=0$). In both cases, the $T_d$ was set up on the average of the hive temperatures when the analysed hive was in good condition, and the $\Delta T$ was chosen after solving for the equation~(\ref{Eq:model}), note that for any  $\Pi>2.66$ the $\Delta T=$ 10°C represents a closer match to an ideal condition.

\subsubsection{Method 2: cross correlations}

\begin{figure*} [hbt!]
\includegraphics[scale=0.6,trim=-3cm 1cm 0cm 0cm, angle =0 ]{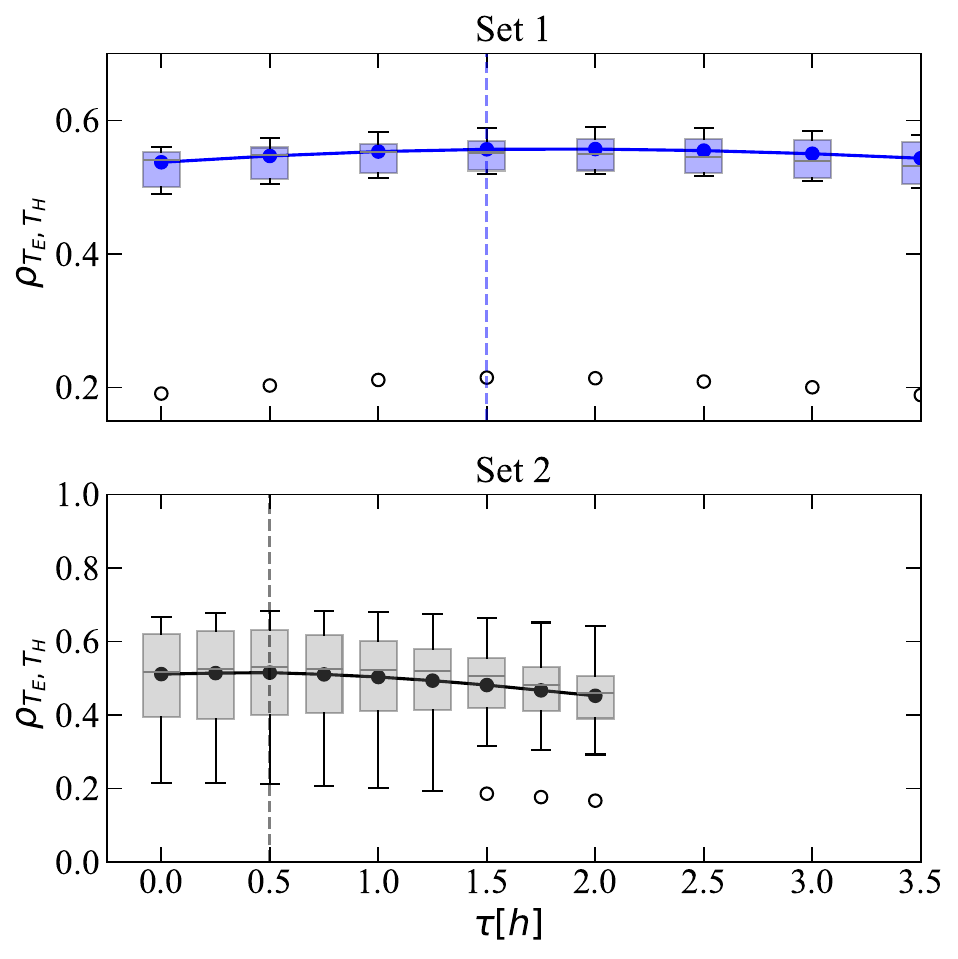}
\caption{The maximum cross-correlation between the hive temperature $T_H$ and environmental temperature $T_E$ happens after a delay time $\tau$ of around $1h$. Each panel shows the dependence of the cross-correlation in equation~(\ref{eq:cross-correlation}) as a function of $\tau$ computed over a set of hives (the box plots mark the median and quartiles, while the dots $\circ$ indicate the outlier points.). The top panel shows the results for our Set 1 of hives, with a maximum around $\tau=1.5 h$. The bottom panel shows the results for our Set 2 of hives, with a maximum around $\tau=0.5 h$. Set 1 consists of 6 hives, and Set 2 consists of 16 hives; the sets were formed based on the time of data collection; refer to supplementary material~(\ref{sec.data}) for detailed information.}
\label{fig.CrossCorrelation}
\end{figure*}

Instead of looking for peak in $T_H$ after an extreme in $T_E$, in our second method we explore the connection between $T_H$ and $T_E$ in the  entire datasets of the hive and environmental temperatures by measure the cross-correlation between the time series. The cross-correlation  between the hive temperature at time $t+\tau$ and the environmental temperature at time $t$ is given by \cite{asuero2006correlation}
\begin{equation}\label{eq:cross-correlation}
   \rho(\tau)= \dfrac{\gamma( \tau)}{\sigma_{T_E}\sigma_{T_H}}=\dfrac{\mathlarger{\mathlarger{\sum}}_{t=1}^{N} \left[ (T_E(t)-\overline{T_E}) (T_H(t+\tau)-\overline{T_H})\right]} {\sqrt{\mathlarger{\mathlarger{\sum}}_{t=1}^{N}(T_E(t)-\overline{T_E})^{2} {\mathlarger{\sum}}_{t=1}^{N}(T_H(t+\tau)-\overline{T_H})^2 }} ,
\end{equation}
where $\gamma(\tau)$ is the cross covariance function and $N$ is the total number of points (temperature records in the time window under consideration). Figure~\ref{fig.CrossCorrelation} shows how the cross correlation $\rho(\tau)$ depends on the delay time $\tau$. Confirming our previous discussion of the delayed impact of $T_E$ on $T_H$, we see that $\rho(\tau)$ is positive and peaks with a lag of approximately $1.5$ h for Set 1 and $0.5$ h for Set 2. Details for the hives that comprise each set can be found in supplementary material ~(\ref{sec.data}).

We now connect the cross-correlation to the observables of interest $m$ (and $\Pi=-\log m$) and $\Delta T$ appearing in equation~(\ref{Eq:model}) and used in the first method. This is done by noting that, at a fixed $\tau$, the cross correlation $\rho$ is connected to the least-squared estimator $\hat{m}$ of the slope between the $T_E(t)$ and $T_H(t+\tau)$ data as
\begin{equation}\label{eq.mrho}
m=\rho(\tau)\dfrac{\sigma_{T_H}}{\sigma_{T_E}},
\end{equation}
where $\sigma_T$ is the standard deviation of the corresponding time series of $T$. We fix a small time window of $7$ days around a time $t$ and determine both the slope $m(t)$, therefore  $\Pi(t)$, at each time step with length $t_{max}$ (see supplementary material~(\ref{S_Cross_correlation}) for a more detailed derivation).  This allows us to estimate $\Pi=-\log m$ from $m$ in equation~(\ref{eq.mrho}), which is computed at the delay time $\tau^*$ with maximum correlation, i.e., $\rho = \rho(\tau=\tau^*)$. Similarly, $\Delta T$ is estimated from equation~(\ref{Eq:model}) as $\Delta T=\frac{\bar{T_H}-\bar{T_d}}{m} +\bar{T_d} -\bar{T_E}$, where all terms on the right-hand side are computed (averaged) over the same time window.

\newpage
\section{Results}

In this section, we test and apply the Methodology introduced in section~\ref{Sec:Methodology} to data of $22$ hives grouped into two sets. Set 1 consists of 6 hives from November 2016 to September 2017, while Set 2 consists of 16 hives from January to March 2020. For more information, see supplementary material~(\ref{sec.data}) for a description of the data and supplementary material~(\ref{S_Pre-processing}) for pre-processing details. The data and codes used in our analysis are available in our repository ~\cite{myrepo2025}.\par
\begin{figure}[htbp]
    \centering
    \begin{minipage}[t]{0.465\textwidth}
        \centering
        \hspace{1.3cm}\textbf{\normalsize Stable Hive Results}\\[0.3cm]
        \includegraphics[scale=0.43, trim=0cm 0cm 0cm 0.0cm, angle=0]{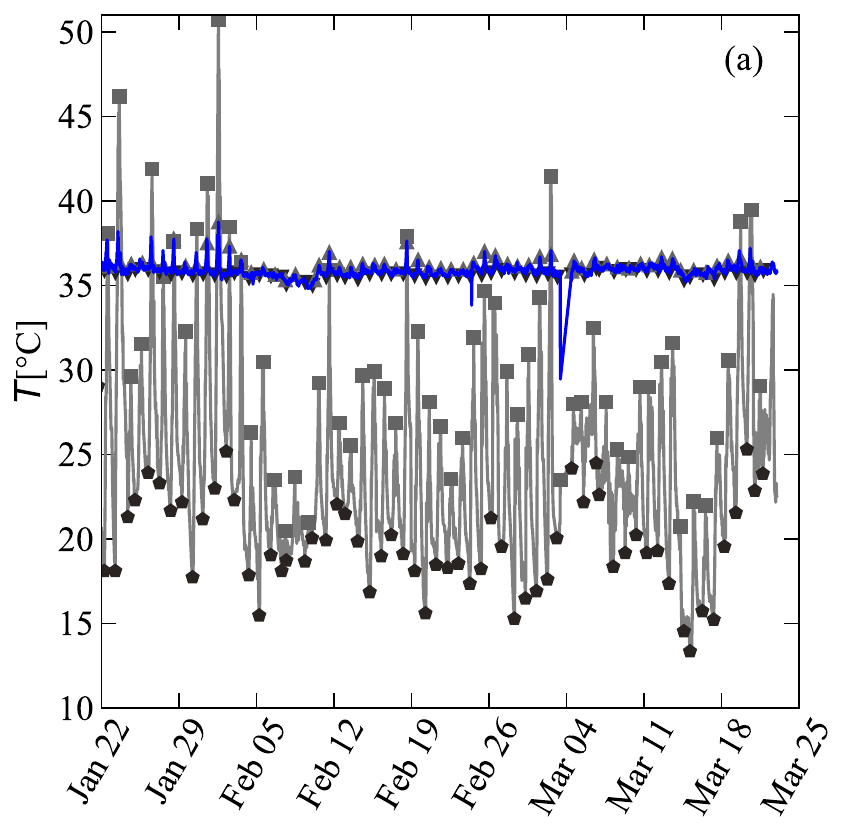}
        \includegraphics[scale=0.43, trim=-0.4cm 0cm 0cm 0cm, angle=0]{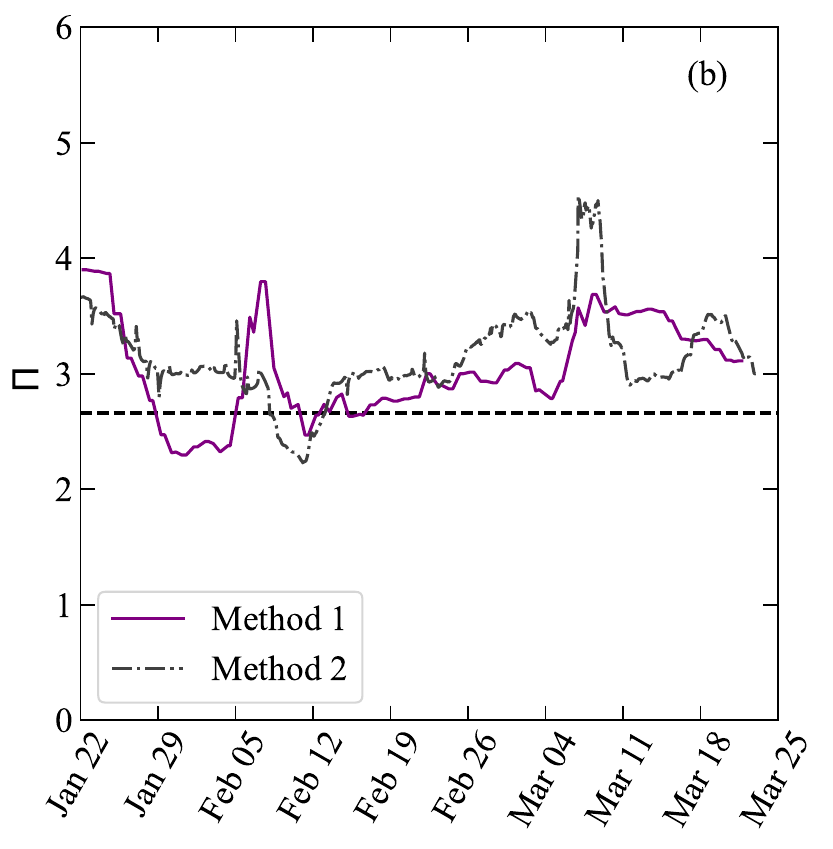}
        \includegraphics[scale=0.43, trim=0.0cm 0cm 0cm 0cm, angle=0]{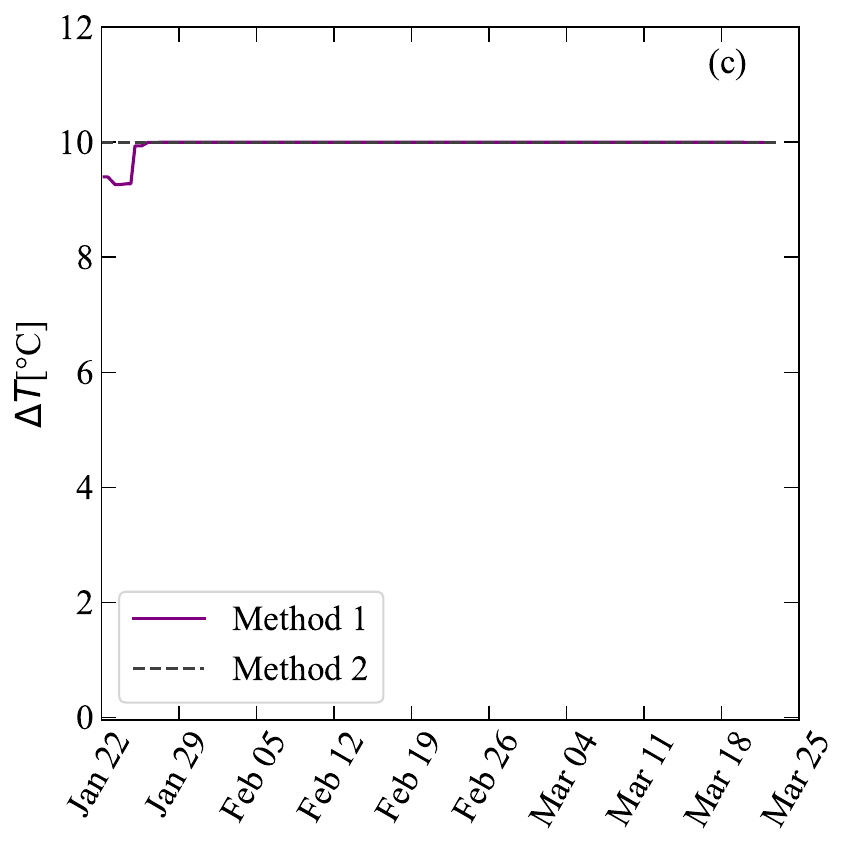}       
    \end{minipage}
    \hspace{.5 cm}
    \begin{minipage}[t]{0.45\textwidth}
        \centering
        \hspace{0.5cm}\textbf{\normalsize Unstable Hive Results} \\[0.3cm]     
        \includegraphics[scale=0.43, trim=0.4cm 0cm 0cm 0.0cm, angle=0]{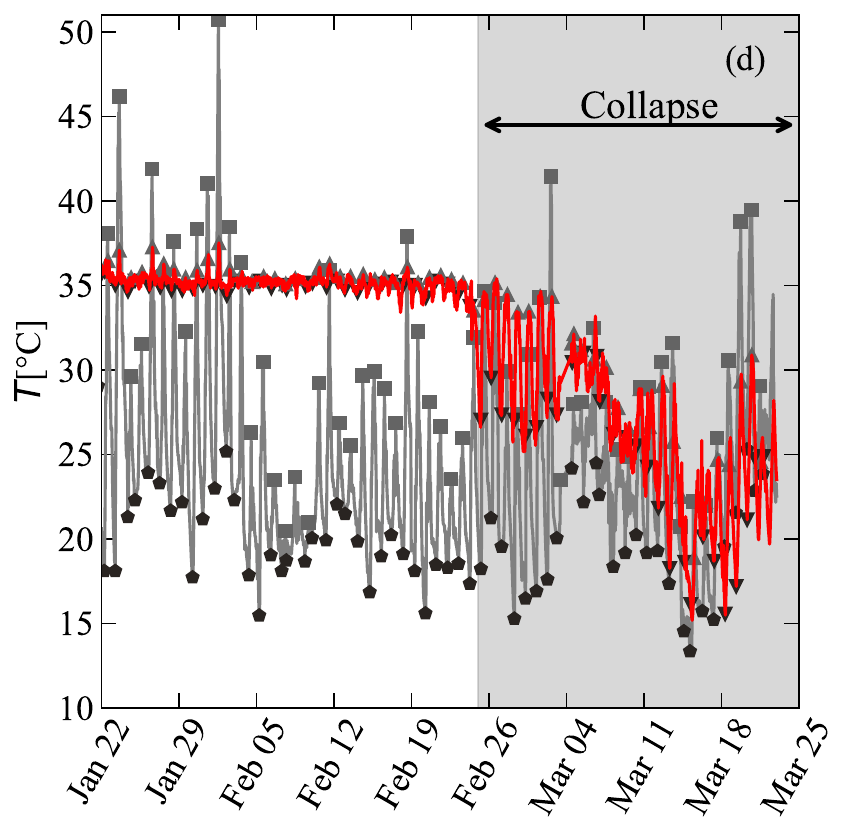}
        \includegraphics[scale=0.43, trim=0cm 0cm 0cm 0cm, angle=0]{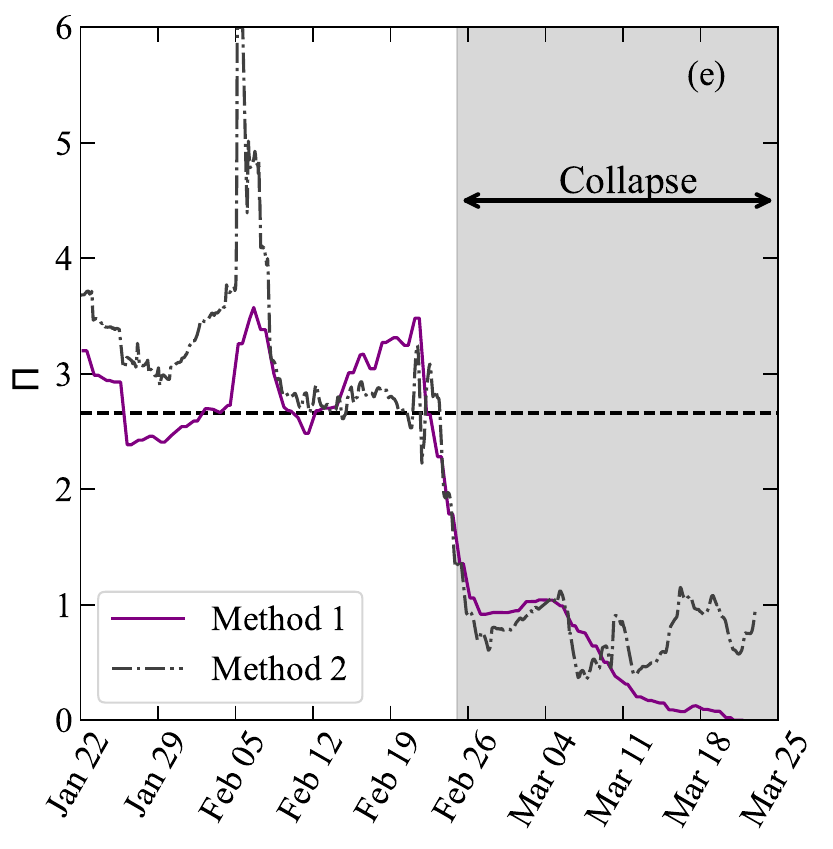}
        \includegraphics[scale=0.43, trim=0.4cm 0cm 0cm 0cm, angle=0]{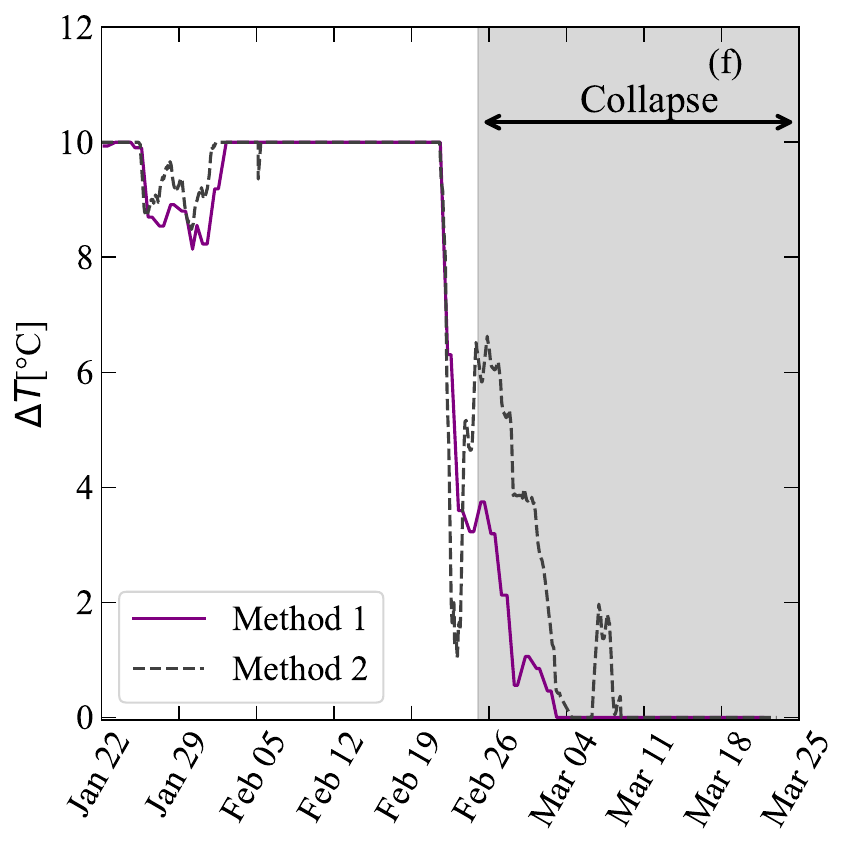}
    \end{minipage}
    \caption{Evolution of $\Pi$ and $\Delta T$ values across two months using Methods 1 and 2 recalling the stable and unstable hives from Figure \ref{Th_Te}. The variations of $\Pi$ and $\Delta T$ reflect the collapse over time for the unstable hive. The analysis was made considering a rolling time window with a span of seven days. The gray shade represents the collapsed regime where the hive lost control of regulating their internal temperature.}
    \label{Fig:Model_results}
\end{figure}

We start by comparing how our two methods perform in a stable hive and in a hive that eventually collapses. The results in figure~\ref{Fig:Model_results} shows that both methods lead to similar estimations of $\Pi(t)$ and $\Delta T(t)$,  succeeding in  differentiating between the two hives, and showing the decay in the status of the fragile hive ahead of the collapse. The results of the two hives were computed during the summer (brood-rearing season), under the same spatial-temporal conditions, and show that one hive cannot regulate its internal temperature (subfigures \ref{Fig:Model_results} d-f) compared to the stable hive (subfigures \ref{Fig:Model_results} a-c), even for warmer temperatures. This plot shows both methodologies can determine if a hive is unstable or performing at a lower level than others.  The gray shaded region marks the collapse period, defined as the time when the mean hive temperature consistently deviates from the optimal range and the standard deviation scales with the environmental temperature (supplementary material~\ref{Sec:Statistical Features}).  The onset of this collapse occurs when $\Pi \approx 1.5$ for both cases.

\begin{figure}[htbp!]
    \centering
    \begin{minipage}[t]{0.45\textwidth}
        \centering
        \hspace{.7cm}\textbf{\normalsize Method 1}\\[0.3cm]
        \includegraphics[scale=0.40,trim=0cm 0cm 0cm 0.0cm, angle =0 ]{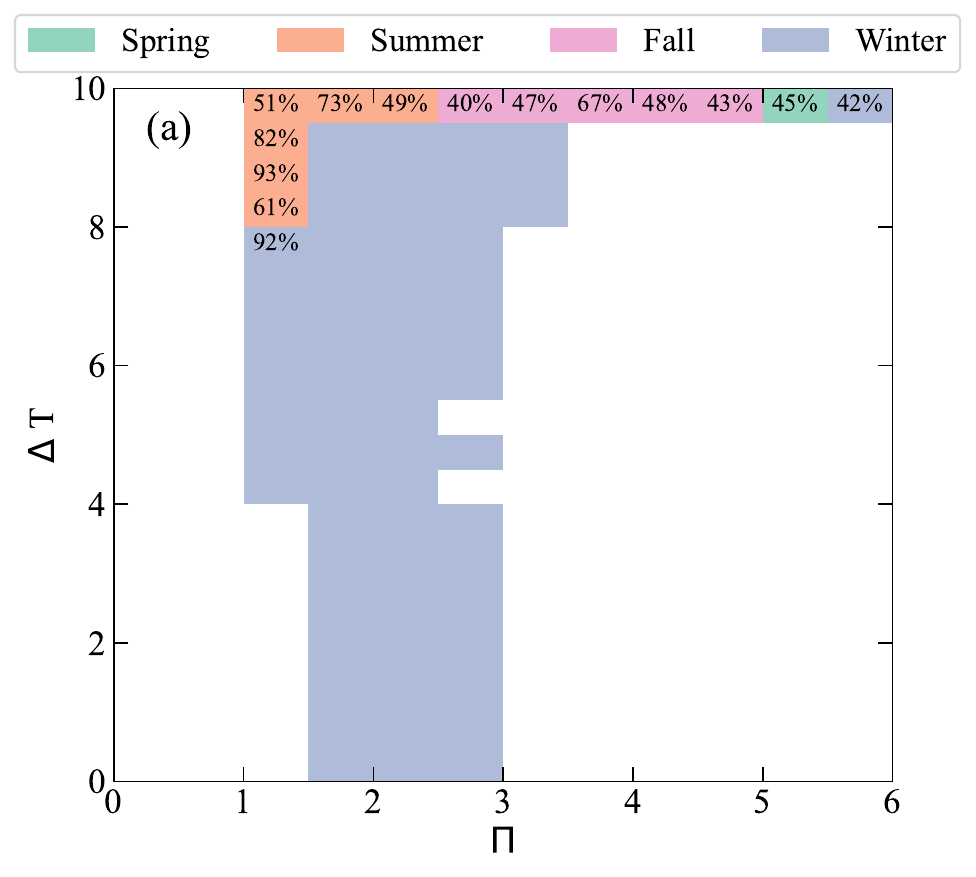}
        \includegraphics[scale=0.40,trim=0cm 0cm 0cm 0.0cm, angle =0 ]{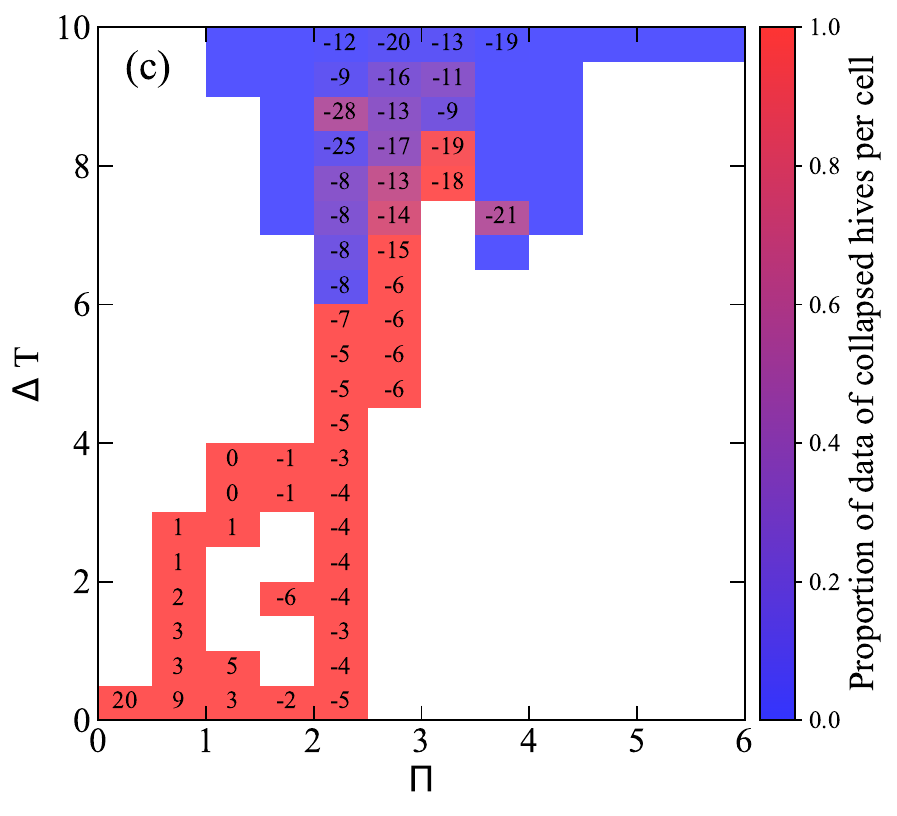}     
    \end{minipage}
    \hspace{1.0 cm}
    \begin{minipage}[t]{0.45\textwidth}
        \centering
        \hspace{.7cm}\textbf{\normalsize Method 2} \\[0.3cm]     
        \includegraphics[scale=0.40,trim=0cm 0cm 0cm 0.0cm, angle =0 ]{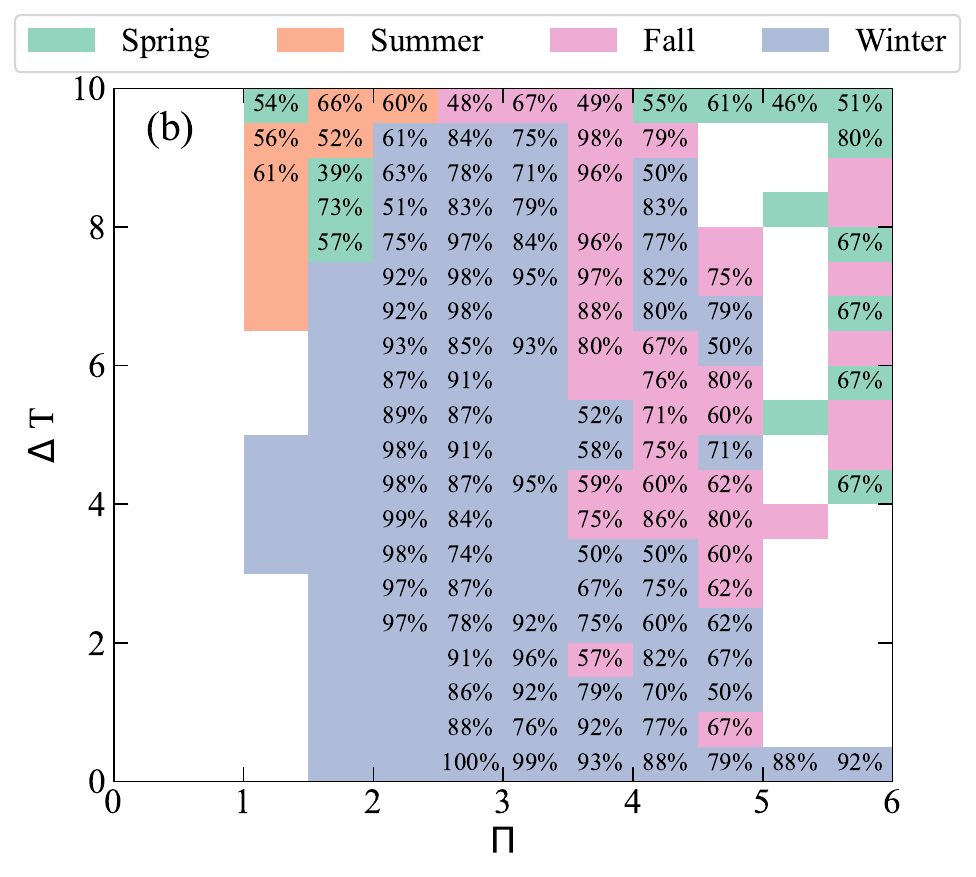}
        \includegraphics[scale=0.40,trim=0cm 0cm 0cm 0.0cm, angle =0 ]{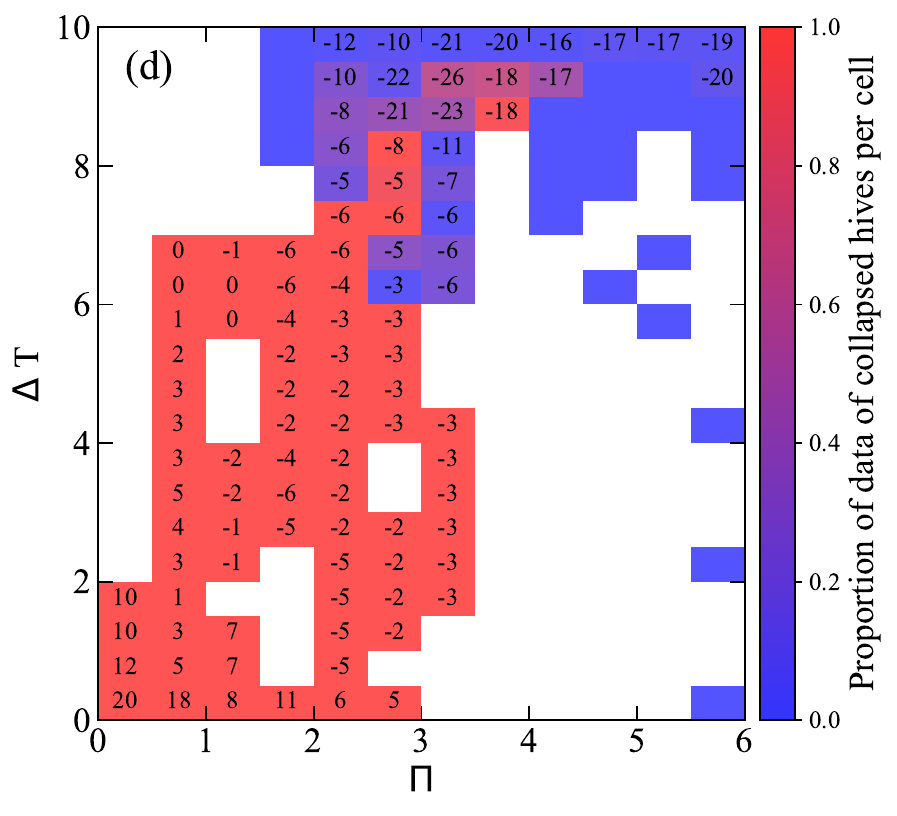}
    \end{minipage}
    \begin{minipage}[t]{0.65\textwidth}
        \hspace{1.cm}\textbf{\normalsize Summary from Methods 1 and 2} \\[0.3cm]     
        \includegraphics[scale=0.40,trim=0cm 0cm 0cm 0.0cm, angle =0 ]{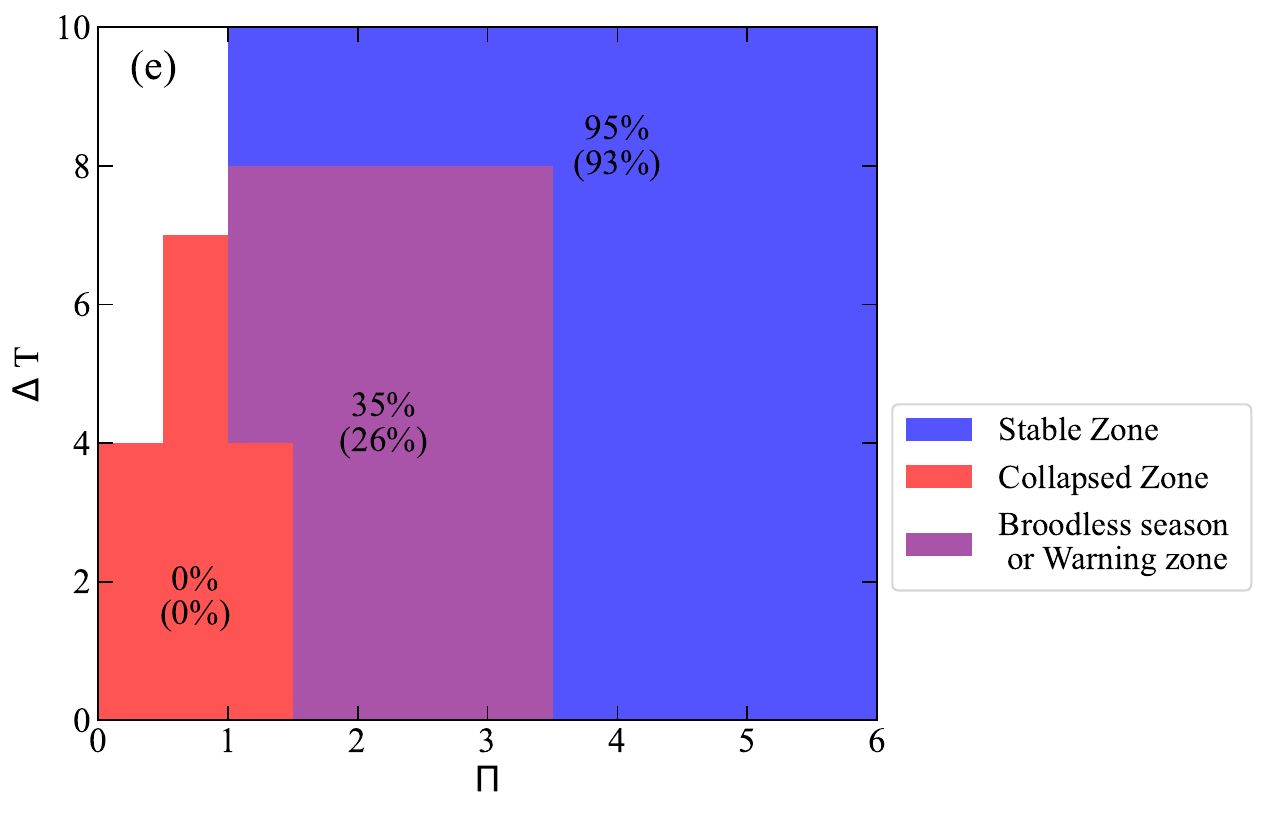}
    \end{minipage}
    \caption{Classification of the status of bee hives based on methods 1 (left plots) and 2 (right plots). (a-b) Summary of the results of 6 hives over approximately a year, where none of the hives collapsed. If data from multiple seasons converge in the same cell, it is coloured based on the season with the highest frequency, expressed as a percentage. (c-d) Relationship between $\Pi$ and $\Delta T$ for 16 hives, where 3 of them eventually collapsed. The numbers indicate the days between the current state and the eventual collapse. (e) Summary for identifying the stable, collapse zones, and their transitions for Methods 1 and 2. The percentage represents the ratio of healthy points within each of the three zones. See supplementary material~(\ref{Grid_computation}) for details.}
    \label{fig_summary}
\end{figure}

We extend the analysis and report the results obtained from our two independent data sets (22 hives) for Methods 1 and 2, respectively. The subfigures \ref{fig_summary} a-b display the analysis of the earlier dataset, Set 1 (6 hives), where no hive failures were identified over approximately a year of study. The outcome for Set 2 (16 hives) is shown in subfigures \ref{fig_summary} c-d and covers summer and early fall of the first quarter of 2020, including the transition of three stable hives to instability until the ultimate collapse. After applying Method 1 and 2 to the 22 hives, we establish that $\Pi<3.5$  and $\Delta T < 8 $  should raise concern for a closer monitoring of the hive excluding in winter season wherein the values of $\Pi$ and $\Delta T$ drop from the normal average and quickly recover in spring. Notably, no stable hive has a $\Pi< 1.0$ even during winter for both methods.  Subfigure \ref{fig_summary}-e contains a simple summary of the results, and the percentage displays the ratio of healthy points within three different zones --- stable, warning, and collapsed--- from hive temperature data during 2020 for Method 1 and Method 2, respectively. The remaining question is how early can we detect a hive that is on its way to collapse. In this manuscript, we characterise a collapse as the point when the mean hive temperature fully deviates from the optimal range and the standard deviation scales with the environmental temperature. Interestingly, hives that eventually collapsed maintained internal temperatures within the optimal range until a week before collapsing. A more detailed sequence of $\Pi$ and $\Delta T$ can be observed in subfigure \ref{fig_summary} c-d where the numbers above each cell represent the difference of days between the current state and the collapsed state, averaged across three hives (hives E, H and M). \par

\begin{figure}[htbp!]
    \centering
    \begin{minipage}[t]{0.47\textwidth}
        \centering
        \hspace{.3cm}\textbf{\normalsize Method 1}\\[0.3cm]
        \includegraphics[scale=0.355,trim=0cm 0cm 0cm 0cm, angle =0 ]{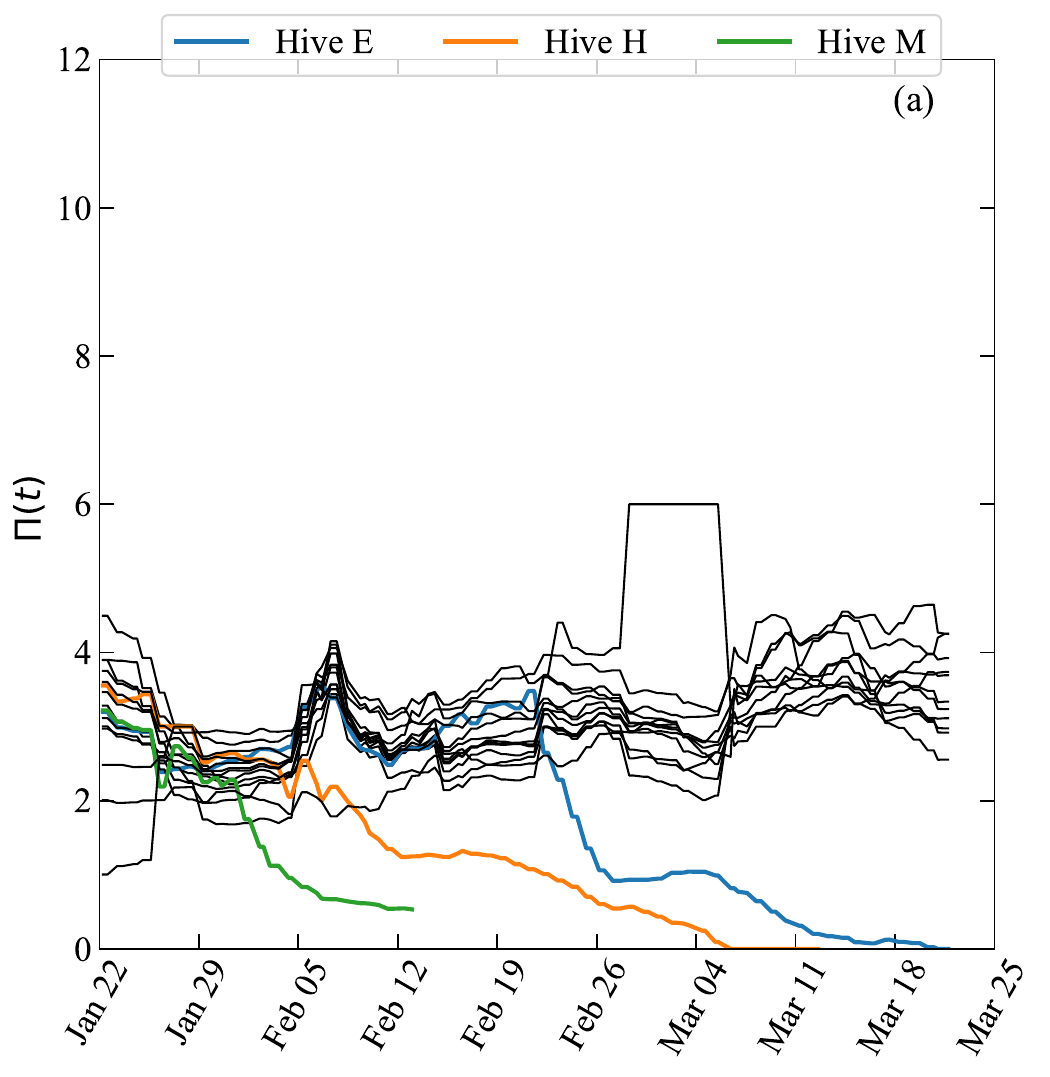}
        \includegraphics[scale=0.355,trim=1cm 0cm 0cm 0cm, angle =0 ]{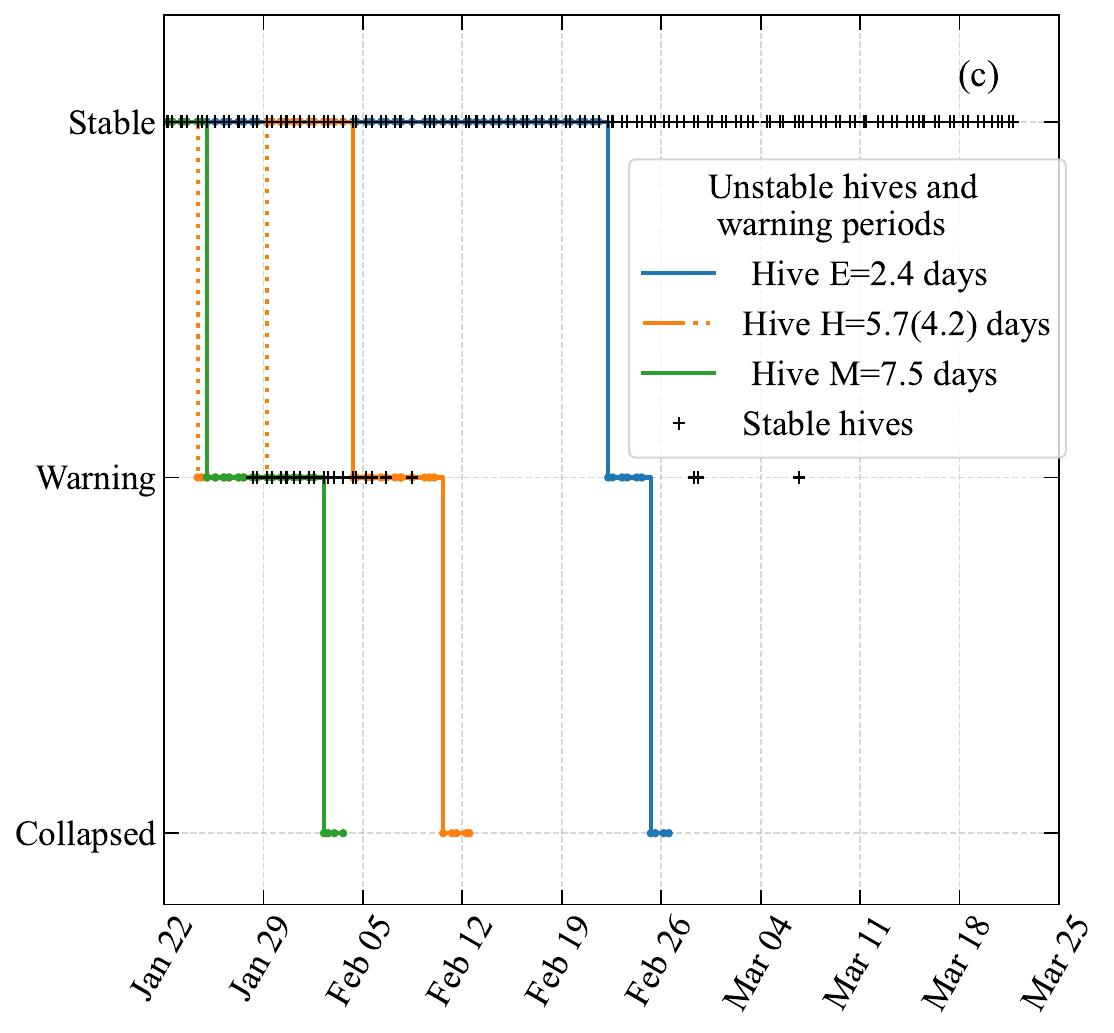}  
    \end{minipage}
    \hspace{.7 cm}
    \begin{minipage}[t]{0.45\textwidth}
        \centering
        \hspace{.3cm}\textbf{\normalsize Method 2} \\[0.3cm]     
        \includegraphics[scale=0.355,trim=2cm 0cm 0cm 0cm, angle =0 ]{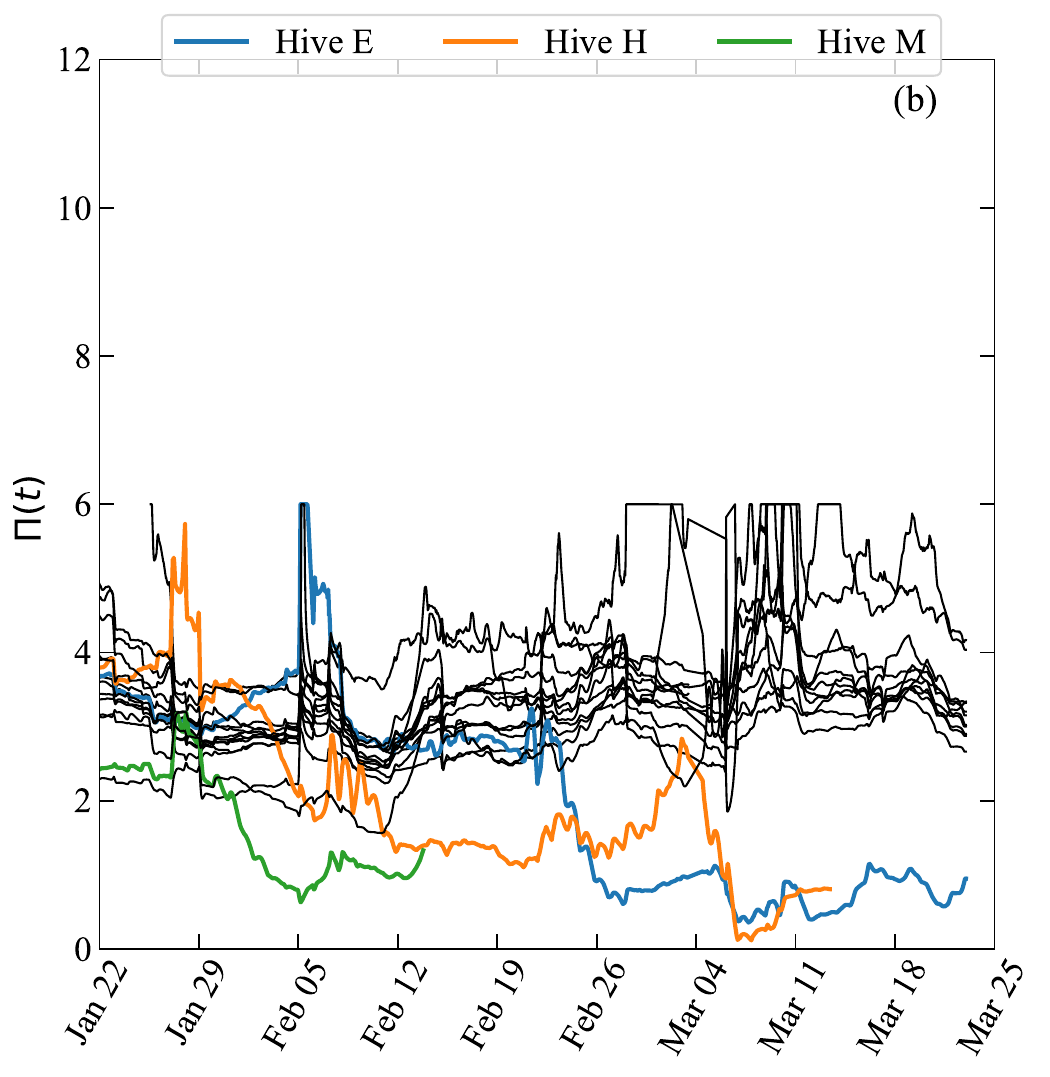}
        \includegraphics[scale=0.355,trim=2.4cm 0cm 0cm 0cm, angle =0 ]{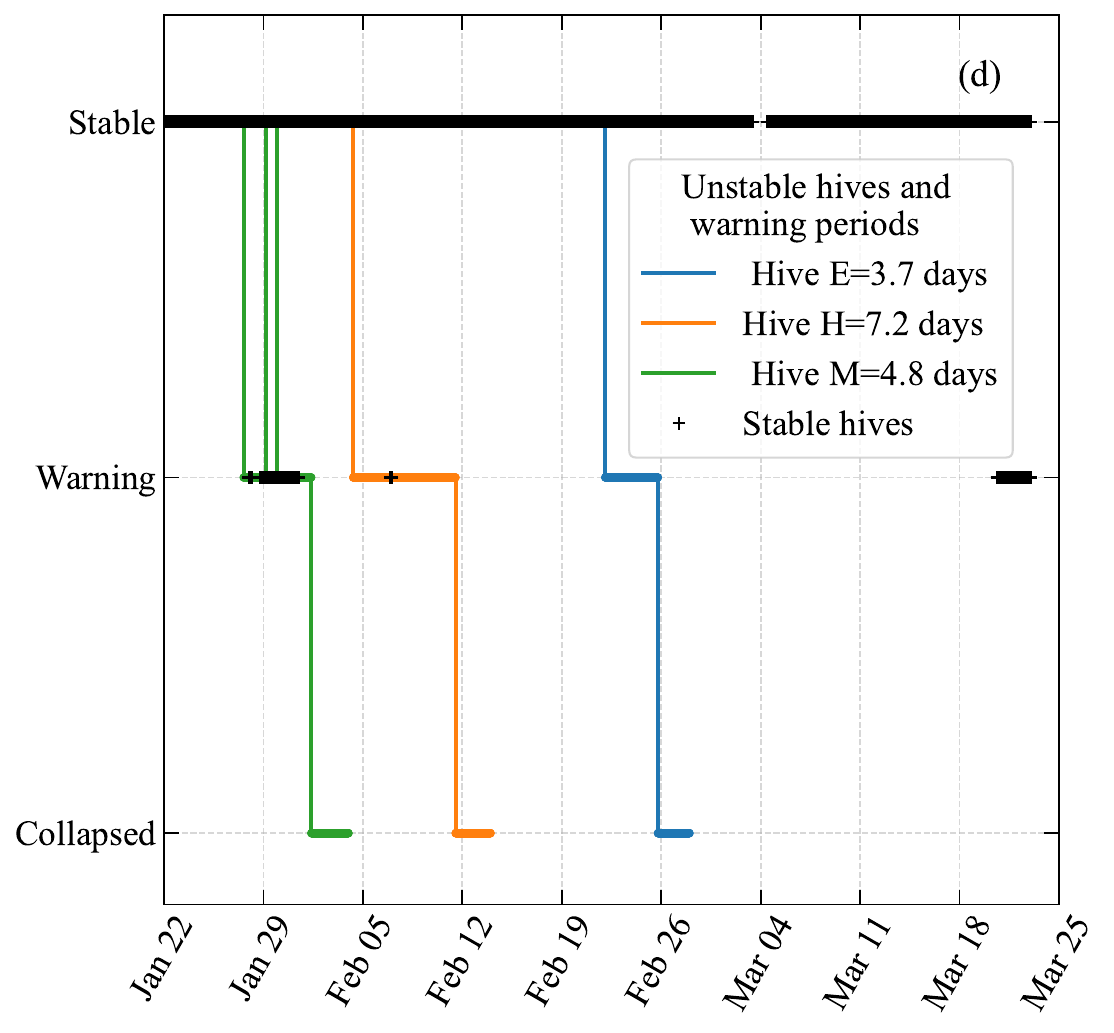}
    \end{minipage}
    \caption{Temporal dependence of the status of 16 hives, all from  Set-2 (supplementary material~(\ref{sec.data})), including the three hives that eventually collapsed (E, H, and M) in a different colour. (a) Value of the indicator $\Pi$ computed with Method 1; (b) Value of the indicator $\Pi$ computed with Method 2. (c-d) Change in status of the hives for Method 1 (c) and 2 (d), respectively. A warning period before the collapse of hives is clearly visible.}
    \label{fig:comparisson}
\end{figure}

We characterised the time-dependent status of each hive (stable, warning, and collapsed) in Set 2 for Methods 1 and 2 respectively (figure \ref{fig:comparisson}). A hive is labelled collapsed if, for two consecutive days, the records consistently agree on that classification. For Method 1 (figure \ref{fig:comparisson}-c), we found that two of the three hives that eventually collapsed exhibited a smooth transition from stability to collapse, with warning periods lasting between 2.4 and 7.5 days. In contrast, among the three hives, hive H showed a more complex pattern, with a warning period of 4.2 days, followed by a recovery, and then a final warning period of 5.7 days before an imminent collapse. Under Method 2\footnote{ Negative cross-correlation values are disregarded, as they indicate a negative slope between the environmental temperature and the hive temperature.} (figure \ref{fig:comparisson}-d), the same hive displayed a single warning period lasting 7.2 days. Furthermore, we found that, of the three hives that eventually collapsed, Method 1 displayed an earlier collapsed state in two of them (hive E and hive H). Interestingly, both methods agreed in identifying hive E as having the fastest transition from stable to collapsed, while hives H and M remained in a warning state for at least five days, providing more time for a successful intervention. Altogether, these observations show that both methods successfully characterise the status of the hives and have high potential for identifying hives at risk well ahead of time, enabling interventions that could save the hives.

Regarding the frequency with which a stable hive was identified as being at risk of collapse or in a warning state, both Methods 1 and 2 flagged two stable hives (Hive A and Hive O), classifying them as in a warning state. Under Method 1, Hive A displayed a single warning record on March 6th. In contrast, under Method 2, it was flagged as being in a warning state at an average frequency of every 20 minutes (closer to the highest possible frequency) for 2.4 days, from March 20th to 22nd. On the other hand, under Method 1, hive O was classified as being in a warning state with an average frequency of approximately 12 hours (closer to the highest possible frequency for daily extreme temperatures) over 11 days, from January 28th to February 8th. Later, the same hive displayed warning behaviour for only one day, February 28th. Method 2 yielded a similar result, with hive O being flagged with an average frequency of about 50 minutes over 10 days, starting on January 27th, and overlapping nearly the same initial warning period flagged by Method 1. The subfigures \ref{fig:comparisson} c-d, present these records for Methods 1 and 2 respectively.

\newpage
\section{Conclusion}

We introduced and tested two methods to quantify the status of honeybee hives based on their response to environmental temperature fluctuations, enabling faster and more cost-effective beehive health monitoring. Building on previous studies demonstrating that in-hive bee response thresholds modulate temperature within their hive's brood nest between 33ºC and 36ºC, our study quantifies their ability or inability to control temperature and links this to the status of the hive. The significance of our results is that it assesses hive stability based only on data that can be easily collected (time-series of hives and environmental temperatures), providing a cheaper (and potentially complementary) diagnostic when compared to alternative approaches that considered hive structure\cite{becher2010brood}, genetic variance\cite{jones2004honey}, hive's weight \cite{arias2023modelling}, colony dynamics \cite{schmickl2007hopomo}, or collective thermoregulation \cite{ocko2014collective,mitchell2023honeybee}. 

We applied our approach to the study of $22$ hives, including $3$ hives that collapsed. We found that both our methods succeed in separating healthy from unhealthy hives, motivating us to propose a monitoring scale that alerts beekeepers when hives are at risk. Although Method 1 detects collapse risks or \textit{warning} states a bit earlier by focusing solely on extreme temperatures, Method 2 offers a more detailed data analysis and a more straightforward approach, enabling closer monitoring of the hives.
Considering both methods, we have established that a hive should be flagged for concern if our key quantification measure $\Pi$ is at $\Pi<3.5$ and elevated to collapse status if $\Pi <1.5$, as this is the threshold for their survival. This threshold is valid from spring to autumn. In winter, the warning threshold should be elevated to only for $\Pi<1.5$. This difference correlates with the fact that winter is a broodless period for honeybees, during which temperature maintenance is less critical to the hive. Brood rearing commonly starts in late winter and rises until summer \cite{stalidzans2013temperature,fahrenholz1989thermal,buchler2020summer,knoll2020honey}. 

\par
Our results and repository \cite{myrepo2025} open up a potential opportunity for comparison with other early warning indicators of collapse, such as signs of approaching tipping points. Motivated by the fact that maintaining an optimal temperature during the brood-rearing season depends on both beehive population and brood-rearing rate \cite{chen2023impacts,stalidzans2013temperature}, we expect that the key quantities identified in our method (e.g., $\Pi$ and $\Delta T$) can be used to detect hives approaching tipping-point collapses such as the ones appearing in bee-hive models~\cite{perry2015rapid}.

\vskip.5pc

\enlargethispage{10pt}

\section*{Acknowledgments}
This work was supported by the Australian Research Council (Grant DP190101994 to TL, MM, EGA)



\vskip.5pc




\clearpage
\bibliographystyle{ieeetr}

\appendix
\section*{Supplementary Material \\ Pre-processing Details and Computational Methods}

\setcounter{page}{0}  
\setcounter{section}{0}
\setcounter{subsection}{0}
\renewcommand{\thesection}{S\arabic{section}}
\renewcommand{\thesubsection}{S\arabic{section}.\arabic{subsection}}

\setcounter{figure}{0}
\renewcommand{\thefigure}{S\arabic{figure}}

\renewcommand{\theequation}{s\arabic{equation}}
\setcounter{equation}{0} 

\section{Datasets}\label{sec.data}

The hives were established on the campus of Macquarie University (33° 46' 05.1" S 151° 06' 46.4" E, Macquarie Park, NSW, Australia).
We analysed a time series of honeybee hives from two distinct data sets. Set-1 consists of 6 hives monitored from 18th November 2016 to 28th September 2017 at 30-minute frequency, while Set-2 consists of 16 hives. From Set-2, 11 hives were observed from 21st January 2020 to 22nd March 2020, 2 hives from 21st January 2020 to 3rd March 2020, and 3 hives that eventually collapsed between January 2020 and March 2020. Data from hives in Set-2 was recorded with a frequency of 15 minutes. To avoid incomplete data and potential bias, for Set-1, our methodologies are presented starting from November 20th, two days after the installation of the devices, while for Set-2 our methodologies are presented starting from January 22nd, as the data from January 21st is incomplete. Regarding the data collection, it was collected in situ using  DS1922L data logger from the iButton\textregistered{} device, which has temperature accuracy of $\pm 0.5$°C from $-10$°C to $+65$°C (DS1922L), $\pm 0.5$ °C from $+20$°C to $+75$°C (DS1922T), with Software Corrections and resolution of 0.0625°C.

\section{Cross Correlation and linear association} \label{S_Cross_correlation}
In this section we derive Eq.~(\ref{eq.mrho}). We can write the model in Eq. \ref{Eq:model} as a straight line $y=mx+b$ considering $x=T_E(t)$, $y=T_H(t+\tau)$, and $b=T_d-(T_d-\Delta T)m$ . The least-squared estimator $\hat{m}$ of $m$ computed over $i=1, \ldots, N$ points $(x_i,y_i)=(x(t_i),y(t_i)$ can be written as
\begin{equation}\label{eq.m2}
\hat{m}= \frac{\sum_{i=1} ^{N}  (x_i-\bar{x}) (y_i-\bar{y})}{\sum_{i=1} ^{N} (x_i - \bar{x})^2}
\end{equation}
Applying the definition of cross correlation in Eq.~(\ref{eq:cross-correlation}) to $x(t),y(t)$ and  using Eq.~(\ref{eq.m2}) we obtain 
\begin{equation}
    \hat{m}\frac{\sigma_x}{\sigma_y}=\frac{\sum_{i=1} ^{N}  (x_i-\bar{x}) (y_i-\bar{y})}{\sum_{i=1} ^{N} (x_i - \bar{x})^2} \sqrt{\frac{\sum_{i=1} ^{N} (x_i - \bar{x})^2/ N-1}{\sum_{i=1} ^{N} (y_i - \bar{y})^2/ N-1}}= \frac{\sum_{i=1} ^{N}  (x_i-\bar{x}) (y_i-\bar{y})}{\sqrt{\sum_{i=1} ^{N} (x_i - \bar{x})^2 \sum_{i=1} ^{N} (y_i - \bar{y})^2} }= \rho_{x,y},
\end{equation}
which corresponds to Eq.~(\ref{eq.mrho}) after identifying $\hat{m}$ with $m$.

\section{Pre-processing data}\label{S_Pre-processing}
For Method 1, extreme temperatures, the maximum environmental temperature is calculated by examining temperatures from 6 a.m. to 6 p.m., while minimum environmental temperatures are determined by looking from 5:00 p.m. to 5:00 p.m. of the next day, as sudden temperature drops are observed during the night or morning, depending on the season. Later on, the effect of this environmental temperature is examined in the hive by inspecting the temperature up to 2.0 hours after the maximum or minimum environmental temperature of the day, $0\,\mathrm{h} < \tau < 2.0 \,\mathrm{h}$. Furthermore, when analysing over a rolling time window (spanning more than a day), any maximum environmental temperature recorded within a day that is lower than the day's highest minimum temperature is disregarded in the analysis, as our focus is solely on extreme temperatures. For Method 2, all the environmental and hive temperature data sets are included in the analysis. 
\section{Statistical features in temperature signals}
\label{Sec:Statistical Features}
\begin{figure*} [hbt!]
\includegraphics[scale=0.367,trim=0cm 0cm 0cm 0cm, angle =0 ]{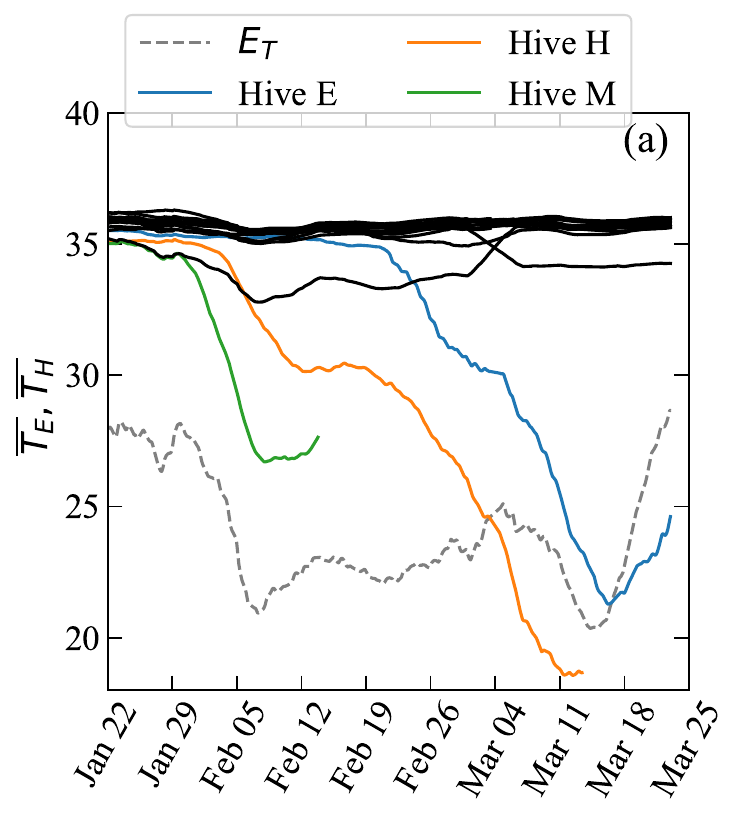}
\includegraphics[scale=0.367,trim=0cm 0cm 0cm 0cm, angle =0 ]{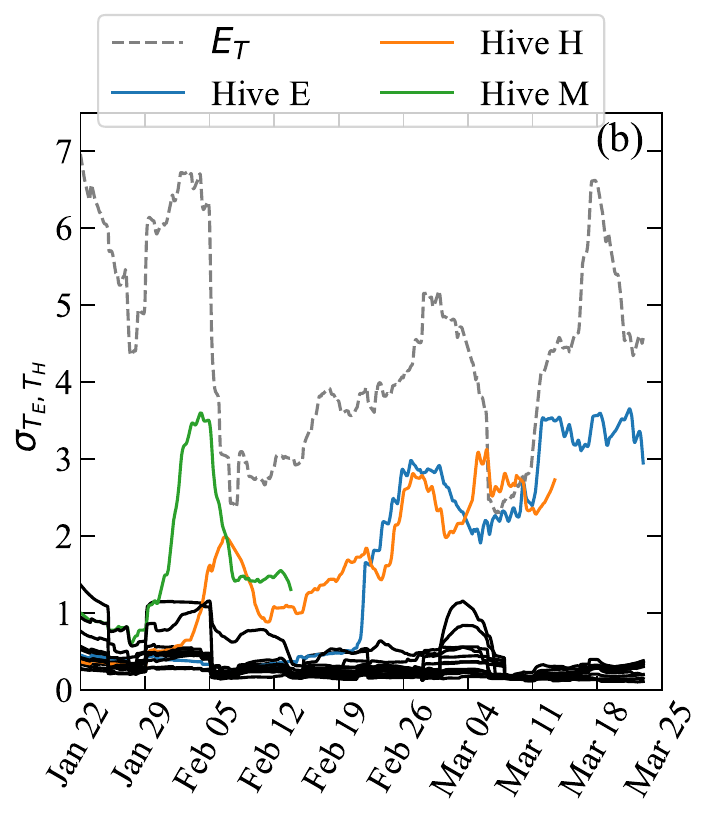}
\includegraphics[scale=0.367,trim=0cm 0cm 0cm 0cm, angle =0 ]{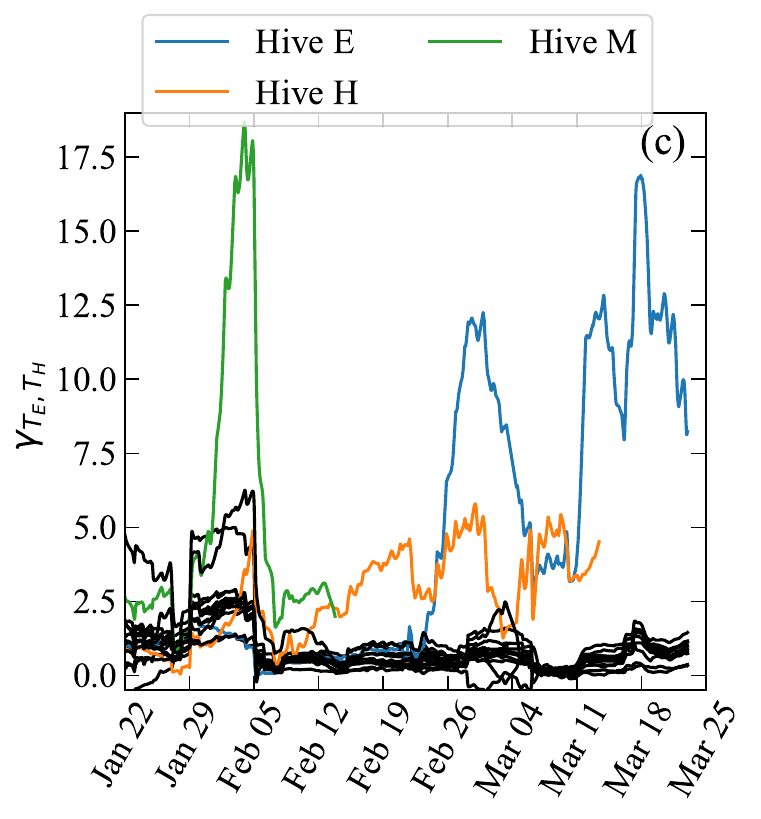}
\caption{Illustration of the time series analysis of the registered environmental temperature and the hive temperature. (a) Calculation of the mean value for each hive $\overline{T_H}$ and for the environmental temperature $\overline{T_E}$. (b) Calculation of the standard deviation for each hive $\sigma_{T_H}$ and for the environmental temperature $\sigma_{T_E}$.  (c) Covariance between the $T_E$ and $T_H$ for Set-2 at $\tau=0.5$ h. In all the cases, the three hives that collapsed, E,H, and M, were distinguished in different colours.}
\label{fig:App1}
\end{figure*}

\begin{figure*} [hbt!]
\includegraphics[scale=0.373,trim=0cm 0cm 0cm 0cm, angle =0 ]{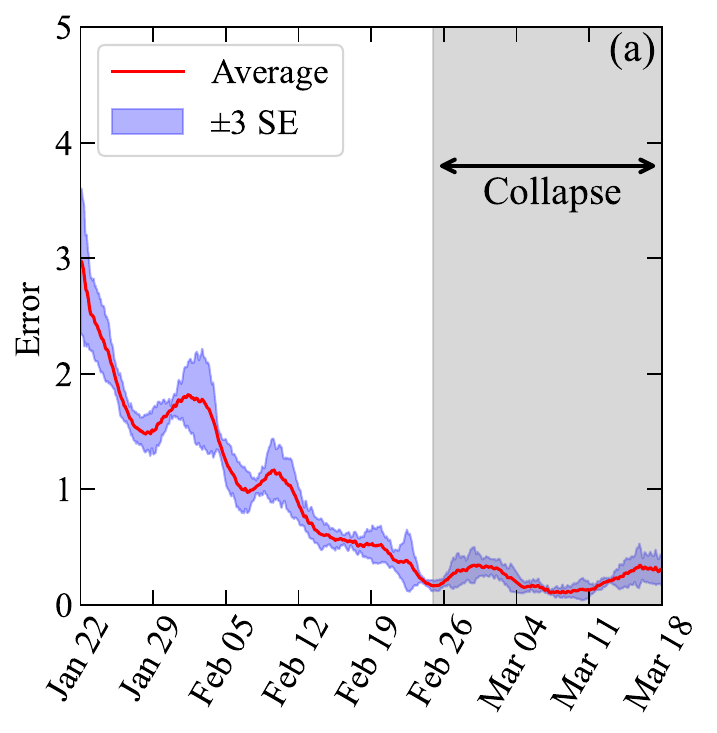}
\includegraphics[scale=0.373,trim=0cm 0cm 0cm 0cm, angle =0 ]{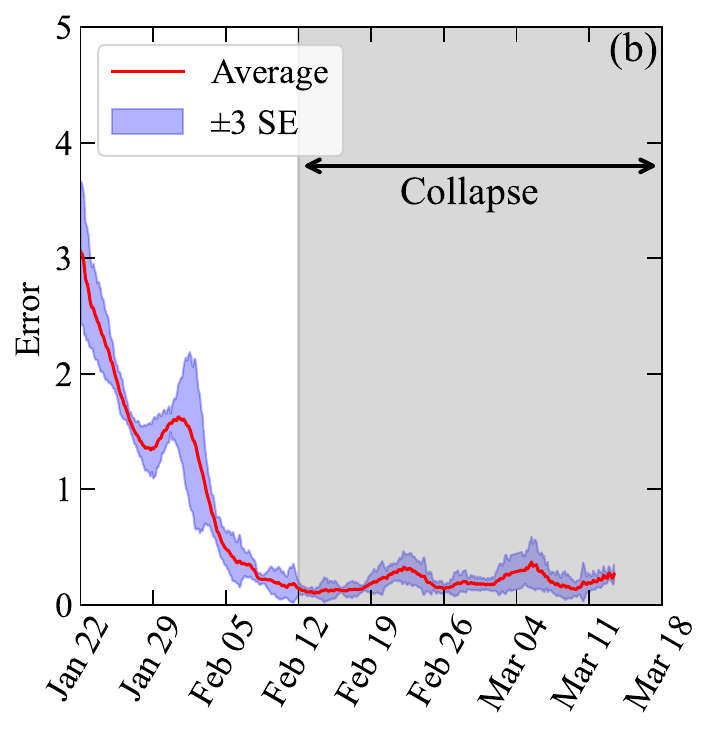}
\includegraphics[scale=0.373,trim=0cm 0cm 0cm 0cm, angle =0 ]{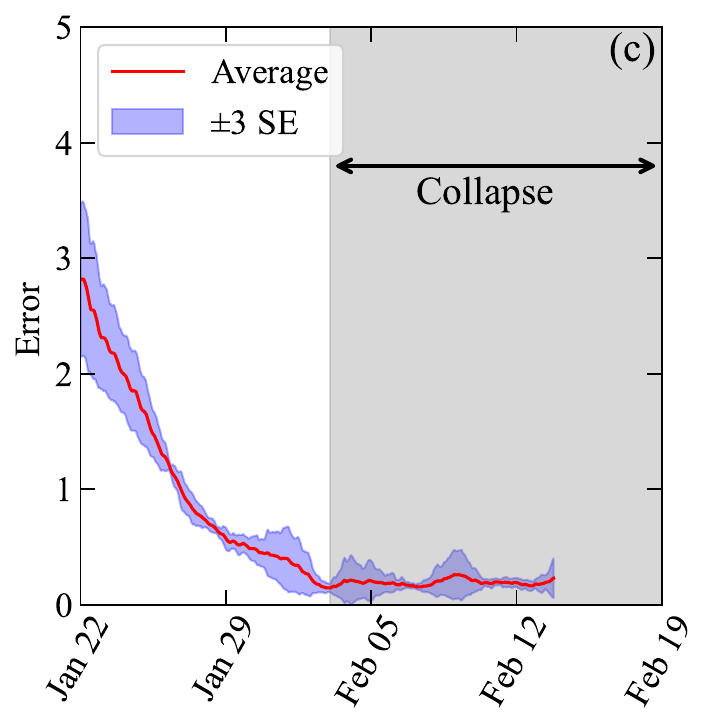}

\caption{Difference between the probability distribution functions of the standard deviation increments for environmental temperature and hive temperature in both stable and unstable hives (a) Hive E, (b) Hive H and (c) Hive M. Error= $\left[ \sum | P(\sigma_{T_E}) - P(\sigma_{T_H}) | \right] ^2$. The average and standard error were calculated for different time windows from 4 to 10 days.}
\label{fig.Error}
\end{figure*}

We characterise a collapse as the point when the mean hive temperature, subfigure \ref{fig:App1}-a , fully deviates from the optimal range and the standard deviation scales with the environmental temperature. Besides observing fluctuations in the standard deviation of collapsed hives similar to those in the environmental temperature, subfigure \ref{fig:App1}-b, we quantify how much these increments differ from those of the environmental temperature. For the three hives that eventually collapsed, we calculate the probability distribution function of standard deviation increments for $T_E$ and $T_H$, respectively, and subtract them to analyse their difference. The squared sum of these absolute differences is presented in figure \ref{fig.Error} as Error. For unstable hives, the error approaches a plateau when the hives lose control of internal temperature regulation. We calculate this error over various time windows ranging from 4 to 10 days.

\section[\texorpdfstring{Computation for the grid $\Pi$ vs $\Delta T$}{Computation for the grid Pi vs Delta T}]
{Computation for the grid $\Pi$ vs $\Delta T$}
\label{Grid_computation}
Subfigure \ref{fig_summary} a-b summarises the results of 6 hives between November 2016 and September 2017 (Set 1), where none of the hives collapsed. The temperature analysis was discretized into 0.5 x 0.5 grids and classified by seasons. The colours of the cells follow this pattern: orange represents summer, pink represents autumn, green represents spring, and sky blue represents winter. If data from multiple seasons converge into the same cell, the cell is coloured based on the season with the highest relative frequency. The displayed percentage indicates the proportion of points corresponding to that dominant season.
For subfigure \ref{fig_summary} c-d the results for 16 hives evaluated between January and March 2020, where 3 of them eventually collapsed. The plotted results depict the relationship between $\Pi$ and $\Delta T$, with both variables, discretized into 0.5 x 0.5 grids to identify potential failure zones. If only failure data converges in one cell, it is coloured in red. If the cell only contains data of healthy hives, it is coloured in blue. If data from both healthy and unhealthy hives converge in a cell, we scaled the colour based on the proportion of data of collapsed hives inside the cell. If there is an absence of data, cells are coloured in white. The colour bars display the proportion of data of collapsed hives per cell when a subfigure contains data from collapsed hives. The numbers over each cell represent the number of days before the collapse occurred in the hives that eventually failed. 
\end{document}